\documentclass[aps,pre,twocolumn,showkeys,superscriptaddress,groupedaddress]{revtex4}
\usepackage{graphicx}
\usepackage{epsfig}
\usepackage{amssymb}
\usepackage{amsmath}
\usepackage{color}
\begin{document}

\title{Randomness is valid at large numbers}

\author{Yeseul \surname{Kim}}
\affiliation{Soft Matter Physics Laboratory, School of Advanced Materials Science and Engineering, SKKU Advanced Institute of Nanotechnology (SAINT), Sungkyunkwan University, Suwon 16419, South Korea}
\author{Byung Mook \surname{Weon}}
\email{bmweon@skku.edu}
\affiliation{Soft Matter Physics Laboratory, School of Advanced Materials Science and Engineering, SKKU Advanced Institute of Nanotechnology (SAINT), Sungkyunkwan University, Suwon 16419, South Korea}
\affiliation{Research Center for Advanced Materials Technology, Sungkyunkwan University,
Suwon 16419, South Korea}
\affiliation{Department of Biomedical Engineering, Johns Hopkins University, Baltimore, Maryland 21218, USA}

\date{\today}

\begin{abstract}
Randomness is a central concept to statistics and physics. Here, a statistical analysis shows experimental evidence that tossing coins and finding last digits of prime numbers are identical regarding statistics for equally likely outcomes. This analysis explains why randomness in equally likely outcomes can be valid only at large numbers.
\end{abstract}


\maketitle

\begin{figure}
\includegraphics[width=8cm]{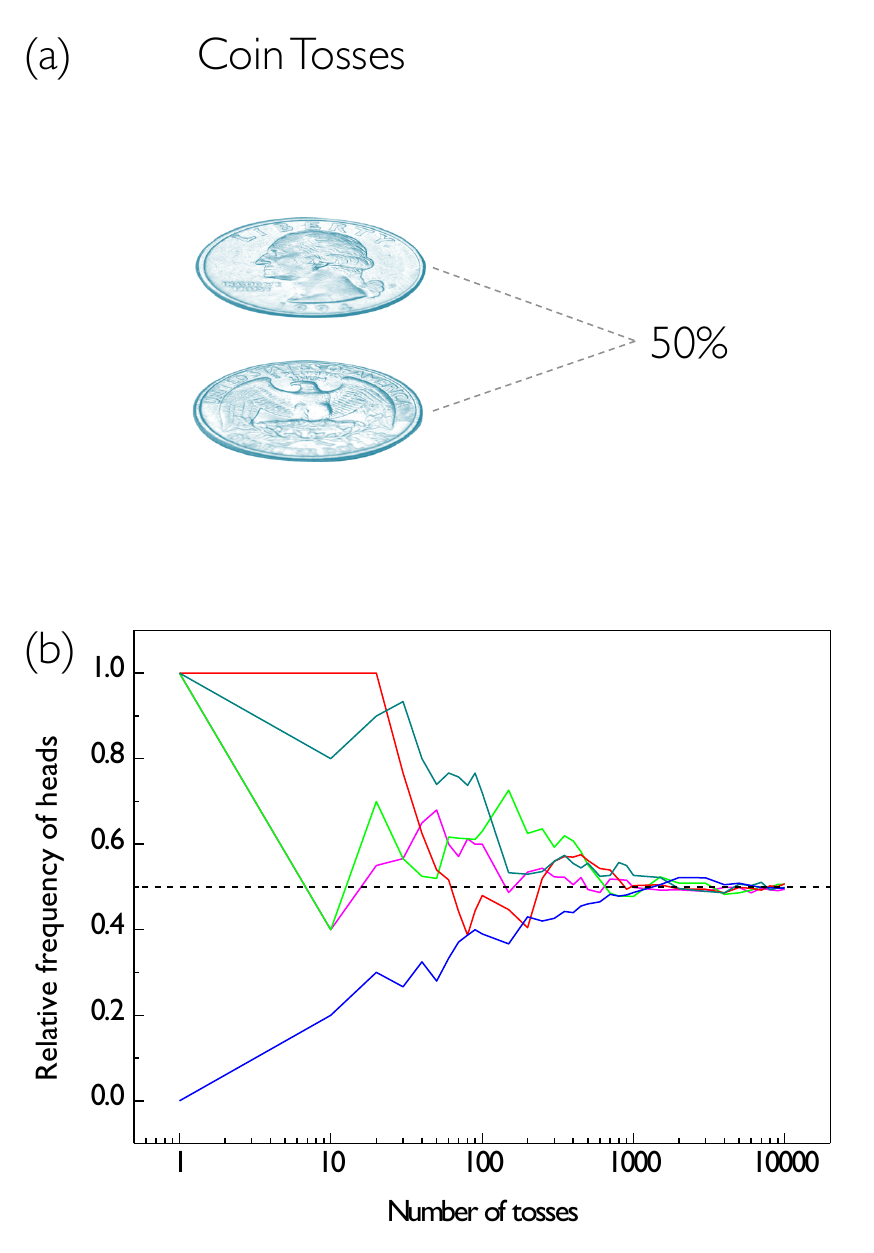}
\caption{Coin tosses. (a) Schematic illustration of a fair coin with two equally likely outcomes (heads or tails): two outcomes equally have 50$\%$ in probability. (b) The relative frequency of heads taken from five experiments (tossing each coin up to $10^{4}$ repetitions). Different experiments are illustrated by different colors. The relative frequency of heads gradually approaches to 50$\%$ [toward the dashed line]. The raw data for five experiments are summarized in the supplementary tables S1$\sim$S5.}
\label{fig:1}
\end{figure}

Randomness is essential in statistics as well as in making a fair decision \cite{Ford,Diaconis,Strzalko08,Mahadevan} and in making pseudo-random numbers \cite{Falcioni,Murphy}. Coin tossing is a basic example of a random phenomenon \cite{Diaconis}: by flipping a coin, one believes to choose one randomly between heads and tails. Coin tossing is a simple and fair way of deciding between two arbitrary options \cite{Strzalko08}. It is commonly assumed that coin tossing is random. For a fair coin, the probability of heads and tails is equal, i.e., Prob(heads) = Prob(tails) = 50$\%$ as illustrated in Fig.~\ref{fig:1}. This situation is valid only under a condition that all possible orientations of the coin are equally likely \cite{Mahadevan}. In fact, real coins spin in three dimensions and have finite thickness, so that coin tossing is a physical phenomenon governed by Newtonian mechanics \cite{Ford,Diaconis,Strzalko08,Mahadevan}. Making a choice by flipping a coin is still important in quantum mechanical statistics \cite{Murphy,Ferrie}. The randomness in coin tossing or rolling dice is of great interest in physics and statistics \cite{O’Hagan,Hayes,Vulovic,Nagler,Strzalko10,Kapitaniak}: coin or dice tossing is commonly believed to be random but can be chaotic in real world \cite{Bellac}.

\begin{figure}
\includegraphics[width=8cm]{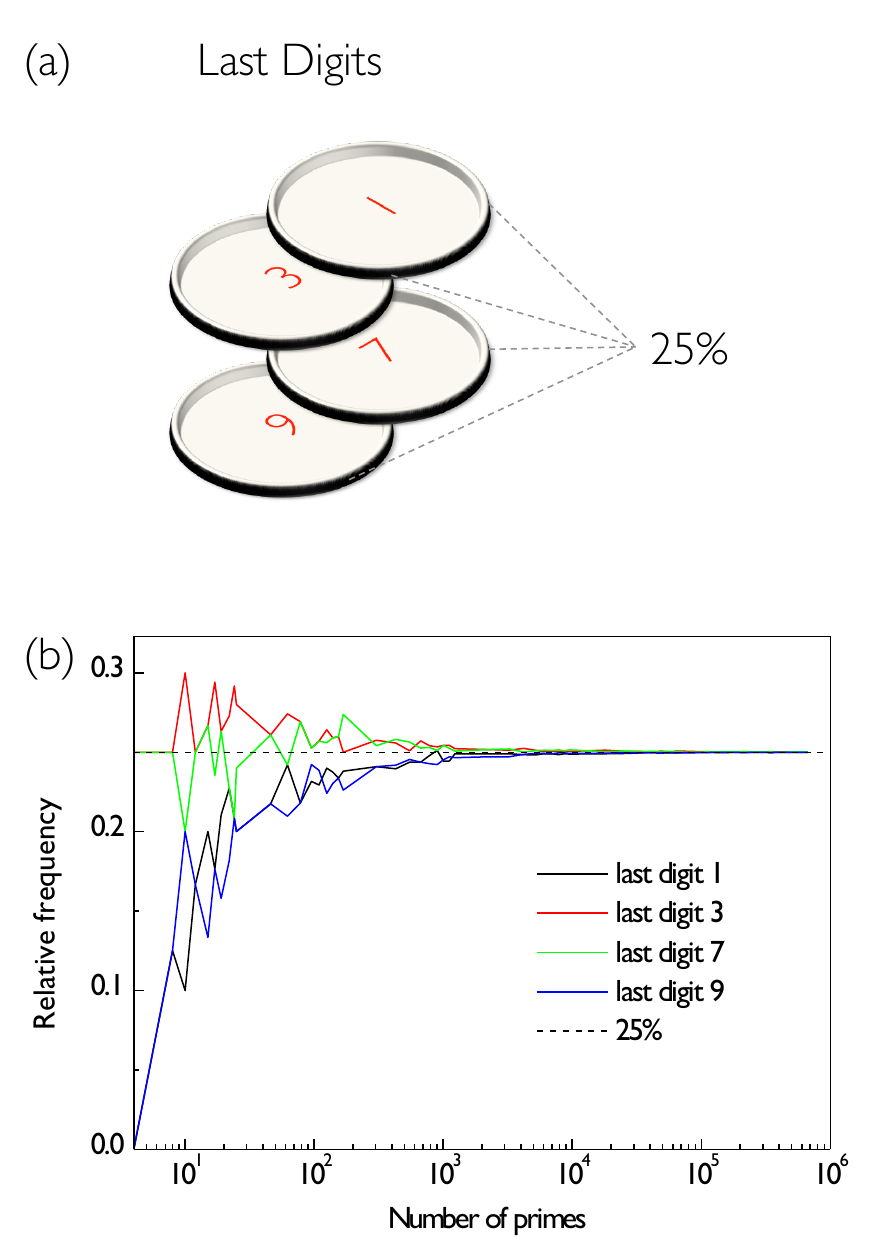}
\caption{Last digits of prime numbers. (a) Schematic illustration of last digits (1, 3, 7, and 9) of prime numbers up to $10^{7}$. The probability of each last digit is expected to be equal as 25$\%$ at large numbers. (b) The relative frequency of last digits gradually approaches to 25$\%$ [toward the dashed line]. The raw data are summarized in the supplementary table S6.}
\label{fig:2}
\end{figure}

A similar situation appears in distribution of prime numbers. Prime numbers are positive integers larger than 1: they are dividable only by 1 and themselves. All primes except 2 and 5 should end in a last digit ($j$) of 1, 3, 7, or 9. In mathematics, the last digits are believed (without a proof) to be random or evenly distributed when numbers are large enough \cite{Granville}. If the last digits of prime numbers come out with the same frequency, then the probability of the four last digits would be equal, i.e., Prob($j$) = 25$\%$ as illustrated in Fig.~\ref{fig:2}. The study of the distribution of prime numbers has fascinated mathematicians and physicists for many centuries \cite{Granville,Pain,Luque,Shao,Tao}. The distribution of prime numbers is essential to mathematics as well as physics and biology. Particularly in many disparate natural datasets and mathematical sequences, the leading digit ($d$) is not uniformly distributed, but instead has a biased probability as $P(d)=\log_{10}(1+1/d)$ with $d = 1, 2, ..., 9$, known as the Benford’s law \cite{Pain,Luque,Shao}. The distribution of last digits of prime numbers is another important topic: in particular, it is unclear that four last digits are random or evenly distributed when numbers are large enough.

\begin{figure}
\includegraphics[width=8cm]{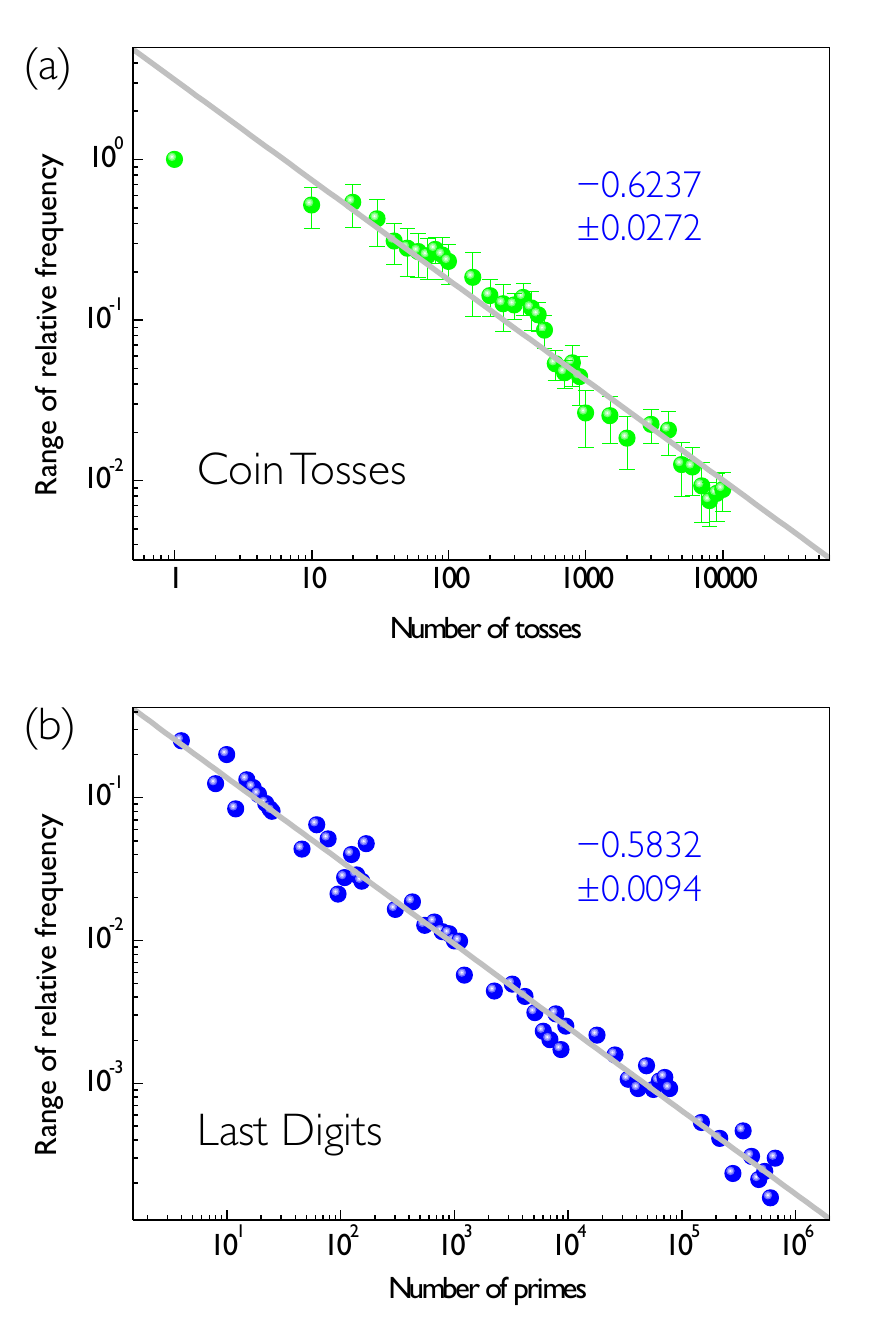}
\caption{Analogy between coin tosses and last digits of primes. The range of relative frequency ($R/N$) (a) for coin tosses between heads (with error bars coming from one standard deviations from five experiments) and tails and (b) for last digits among four last digits (with no error bars because the prime numbers are deterministic). In both cases, a power-law scaling of $R/N \sim N^{\beta}$ is found with $\beta = {-}0.6237$ (the standard error = $\pm0.0272$) for coin tosses and $\beta = {-}0.5832$ (the standard error = $\pm0.0094$) for last digits. The trend line is taken as $R/N = 3.1461 N^{-0.6237}$ (the adjusted $R^{2} = 0.94085$) in (a) and $R/N = 0.5294 N^{-0.5832}$ (the adjusted $R^{2} = 0.98607$) in (b). This result shows that randomness can be valid only at large numbers for both cases.}
\label{fig:3}
\end{figure}

In this article, we present experimental evidence from a statistical analysis, as highlighted in Fig.~\ref{fig:3}, indicating that tossing coins and finding last digits of primes are intrinsically {\it identical} in statistics with respect to equally likely outcomes. This analysis explains that randomness can be valid only at large numbers.

There are many examples for equally likely outcomes: representatively, coin tossing is believed to occur with a probability of 50$\%$ between heads and tails. For repeated experiments with a same sample, if its frequency between expected outcomes is equal, one can say: the expected outcome of the sample is {\it random}. Here we suggest a simple way to define the randomness concerning equally likely outcomes at large numbers.

The frequency of each outcome ($n_{i}$) can vary complicatedly according to experiments and conditions. The relative frequency of an outcome ($f_{i} = n_{i}/N$) is taken by dividing $n_{i}$ by the total number of repetition ($N$ or equally the size of the sample). The range of frequency ($R$) is defined as the difference between the maximum frequency ($n_{i}^{max}$) and the minimum frequency ($n_{i}^{min}$) and consequently described as $R = (n_{i}^{max} {-} n_{i}^{min})$. In statistics, it is well known that the range ($R$) tends to be larger, the large the size of the sample ($N$) \cite{Hozo,Wan}. This tendency can be described by a power-law scaling as $R \sim N^{\alpha}$ where $0< \alpha < 1$. Such a power-law scaling commonly appears in statistics and physics \cite{Newman,Frank}. Additionally, the range of relative frequency ($R/N$) between equally likely outcomes is defined as $R/N = (f_{i}^{max} {-} f_{i}^{min})$, which is equivalent to $(n_{i}^{max} {-} n_{i}^{min})/N$. From $R \sim N^{\alpha}$, $R/N$ should have a simple relation as $R/N \sim N^{\beta}$, where $\beta = \alpha {-} 1$ (here note that $\beta < 0$ because $\alpha < 1$). The statistical constraint of $R/N \sim N^{\beta}$ ($\beta < 0$) implies that the frequency of each outcome should become equal (because $R/N \rightarrow 0$) as the total number of repletion increases ($N \rightarrow \infty $). Consequently, the condition of $R/N \rightarrow 0$ at $N \rightarrow \infty $ explains why randomness is valid only at large numbers, which is known as the law of large numbers in probability theory. In this study, we would like to identify the $\beta$ exponents for equally likely outcomes; in particular, coin tosses (with two outcomes) and last digits of prime numbers (with four outcomes).

First, we conducted experiments for coin tossing. To rule out physical and mechanical aspects of tossed coins, we used an online virtual coin toss simulation application (http://www.virtualcointoss.com) with an ideal coin of zero thickness, where there is no bias between heads and tails, ensuring the equal probabilities for heads and tails. Our experiments with perfectly thin coins enable us to consider only the statistical features of the coin-tossing problems. We carried out separately five experiments. The frequency of heads ($n_{H}$) or tails ($n_{T}$) for each experiment was recorded with the number of tosses ($N$) (equally the size of the sample). The relative frequencies ($f_{H} = n_{H}/N$ or $f_{T} = n_{T}/N$ for heads or tails), the range of frequency [$R = (n_{i}^{max} {-} n_{i}^{min})$ where $i =$ heads or tails], and the range of relative frequency [$R/N = (n_{i}^{max} {-} n_{i}^{min})/N$] were summarized in the supplementary tables S1$\sim$S5. Each of experiments was illustrated by different colors.

In turn, we examined the last digits of prime numbers. As well known, all prime numbers except 2 and 5 should end in a last digit (1, 3, 7, or 9) and the last digits are expected to be random when numbers are large enough, which suggests that the frequency of four last digits should be equal, i.e., Prob($j$) = 25$\%$ [Fig.~\ref{fig:2}(a)]. For the prime numbers in base 10 for integers up to $10^{7}$ (where totally 664,579 prime numbers exist), we counted the frequency of each last digit ($n_{j}$ where $j =$ 1, 3, 7, or 9), the range of frequency [$R = (n_{j}^{max} {-} n_{j}^{min})$], and the range of relative frequency [$R/N = (n_{j}^{max} {-} n_{j}^{min})/N$], as summarized in the supplementary table S6. Here the number of prime numbers ($N$) (including 2 and 5) is equivalent to the size of the sample. 

The statistical uncertainties were checked for coin tossing experiments in the plot of $R/N$ with $N$ [Fig.~\ref{fig:3}(a)] by measuring one standard deviation from five experiments (from five data points for $R/N$ for a given $N$). However, the prime numbers and the range of relative frequency were completely deterministic for integer numbers up to $10^{7}$, which implies no errors in the plot of $R/N$ with $N$ [Fig.~\ref{fig:3}(b)].

For coin tosses, the relative frequency of heads for five experiments differently varies at small numbers but similarly converges on the expected value (50$\%$) at large numbers [toward the dashed line in Fig.~\ref{fig:1}(b)], which supports that coin tossing is a problem of equally likely outcomes. The well-known statistical feature of that the range ($R$) tends to be larger, the large the size of the sample ($N$) suggests a power-law scaling as $R \sim N^{\alpha}$ ($0 < \alpha < 1$. On this basis, we expected a simple relation for the range of relative frequency for heads and tails, denoted $R/N = (n_{i}^{max} {-} n_{i}^{min})/N$ ($i =$ heads or tails) as $R/N \sim N^{\beta}$ where $\beta = \alpha {-} 1 < 0$. As illustrated in Fig.~\ref{fig:3}(a): $R/N = 3.1461 N^{-0.6237}$ for the trend line, we obtained $\beta = {-}0.6237$ (the standard error = $\pm$0.0272) for five coin tossing experiments (error bars came from one standard deviations). This result clearly supports the validity of Prob($i$) = 50$\%$ by $R/N \rightarrow 0$ at $N \rightarrow \infty$, indicating statistical evidence of randomness for coin tosses at large numbers, which is consist with a common belief about coin tossing \cite{Hayes}.

For last digits of prime numbers, the relative frequency of last digits finally approaches to the ultimately expected value (25$\%$) [toward the dashed line in Fig.~\ref{fig:2}(b)]. The range of frequency among last digits increases with the total number of primes as a power-law scaling of $R \sim N^{\alpha}$ with $\alpha \approx 0.4$, which is similar to the case of coin tossing. The range of relative frequency among last digits, denoted $R/N = (n_{j}^{max} {-} n_{j}^{min})/N$ (where $j =$ 1, 3, 7, or 9), shows $R/N \sim N^{\beta}$ where $\beta = {-}0.5832$ (the standard error = $\pm0.0094$) for last digits [Fig.~\ref{fig:3}(b): $R/N = 0.5294 N^{-0.5832}$ for the trend line], which is identical to the case of coin tossing. This result supports the validity of Prob($j$) = 25$\%$ for one of four last digits by $R/N \rightarrow 0$ at $N \rightarrow \infty$, indicating that the last digit of primes would occur with the same frequency at large numbers.

The above two examples of equally likely outcomes lead to the same results: as the size of the sample ($N$) increases, the range of relative frequency ($R/N$) decreases, following the power law scaling as $R/N \sim N^{\beta}$. Here the $\beta$ exponents were found as approximately ${-}0.6$ for coin tossing experiments with Prob($j$) = 50$\%$ for two outcomes and last digits of primes with Prob($j$) = 25$\%$ for four outcomes. (The difference in the pre-factor is mostly uninteresting in the power-law scaling \cite{Newman}). This result shows that randomness can be valid only at large numbers for both cases. This result provides experimental evidence that tossing coins and finding last digits of prime numbers are intrinsically identical with respect to equally likely outcomes.

In conclusion, we introduced a simple expression for randomness at large numbers. From statistical analyses of coin tosses and last digits of primes, we showed that the range of relative frequency between equally likely outcomes ($R/N$) decreases as the total repetition number ($N$) increases. A power-law scaling for $R/N$ versus $N$ in both cases was found as $R/N \sim N^{\beta}$ ($\beta \approx {-}0.6$), implying that the frequency of each outcome becomes equal ($R/N \rightarrow 0$) as the total number of repletion increases ($N \rightarrow \infty$). The condition of $R/N \rightarrow 0$ at $N \rightarrow \infty$ explains why randomness is valid only at large numbers. This result consequently supports that finding last digits of primes is intrinsically identical to tossing coins in statistics: both cases are the same problems of equally likely outcomes. Finally our finding of the power-law relation between the range of relative frequency among equally likely outcomes and the total number of repetition would be significant to understand the validity of randomness at large numbers (as known as the law of large numbers), which would be important in statistics, physics, and mathematics.

{\bf Acknowledgments.} This research was supported by Basic Science Research Program through the National Research Foundation of Korea (NRF) funded by the Ministry of Education (Grant No. NRF-2016R1D1A1B01007133 and Grant No. 2019R1A6A1A03033215).

\end{document}